\documentclass{elsart} 
 
\usepackage{graphicx} 
\usepackage{float} 
 
\usepackage{citesort} 
 
\usepackage{amssymb} 
 
\def \rb {\raisebox{.05ex}[.2cm][.2cm]} 
 
\begin{document} 
 
\begin{frontmatter} 
 
 
 
\title{Estimation of time delay by coherence analysis} 
 
 
\author[adr1,adr2]{R. B. Govindan}, 
\author[adr1]{J. Raethjen}, 
\author[adr1]{F. Kopper},
 \author[adr2]{J. C. Claussen},
\author[adr1]{G. Deuschl\corauthref{cor1}} 
\corauth[cor1]{Prof. G. Deuschl, Department of Neurology, Christian-Albrechts 
University of Kiel, 
Schittenhelmstrasse 10, D-24105 Kiel, Germany} 
\ead{g.deuschl@neurologie.uni-kiel.de} 
\address[adr1] {Department of Neurology, Christian-Albrechts University of Kiel, 
Schittenhelmstrasse 10, D-24105 Kiel, Germany.}
\address[adr2] {Institut f\"ur Theoretische Physik und Astrophysik, Christian-Albrechts University of Kiel, 
Leibnizstrasse 15, D-24098 Kiel, Germany.} 
\begin{abstract} 
Using coherence analysis (which is an extensively used method to study 
the correlations in frequency domain, between two simultaneously measured 
signals) we estimate 
the time delay between two signals. This method is suitable for time delay
estimation of narrow
band coherence signals for which the conventional methods cannot be reliably applied. 
We show by analysing coupled R\"ossler attractors with a known delay, that the  
method yields satisfactory results. Then, we apply 
this method to human pathologic tremor. The delay between simultaneously measured traces of Electroencephalogram 
(EEG) and Electromyogram (EMG) data of subjects with essential hand tremor is
calculated. We find that there is a delay of 11-27 
milli-seconds ($ms$) 
between the tremor correlated parts (cortex) of the brain (EEG) and the 
trembling hand (EMG) which is in agreement with the experimentally observed delay value of 
15 $ms$ for the cortico-muscular conduction time. By surrogate analysis we 
calculate error-bars of the estimated delay. 
 
\end{abstract} 
 
 
\end{frontmatter} 
 
\section{Introduction} 
The time delay between two dynamical systems can provide information on conduction velocity, and the nature and origin of 
coupling, between the processes. So it is necessary to use 
a well validated method for this purpose. Literature on time delay is vast and 
well documented in \cite{ml1,ml2}. Often, in physiological time series 
analysis, a single method cannot be made unique to be applicable for a wide class of 
data stemming from the processes seemingly operated by similar 
mechanisms. Methods used for  
time delay estimation are no exception from this fact. Here, we use a 
spectral based method, maximising coherence, for the time delay 
estimation \cite{carter}. First, we apply this  
method to uni- and bi-directionally coupled 
R\"ossler attractors with a known delay between the systems. 
 In both cases, the results obtained are in good 
agreement within a narrow range of the delay used in the simulation. Then, we 
apply  
this method to simultaneously recorded traces of Electroencephalogram 
(EEG) and Electromyogram (EMG) of subjects with essential tremor a well known
pathological form of hand tremor. We obtain a delay 
of 11-27 milli seconds ($ms$) between the tremor correlated cortical activity (EEG) and
EMG and the results are reasonably within the range of experimentally observed 
value for cortico-muscular transmission  \cite{roth}. This result, further confirms the involvement of cortex in
the generation of essential tremor \cite{hel1}.  
 
The paper is organised as follows: In section 2, we discuss in detail, the 
methodology of coherence  
analysis. Then, we extend the coherence analysis for time delay 
estimation. Due to a time delay there will be a time
misalignment between the two time series thereby causing a reduction in the
coherence estimated between them. In order to compensate for the reduction in
coherence due to delay, we shift one of the time series keeping
the other constant and estimate the coherence as a function of the
shift. This method has been successfully applied to estimate the time 
delay between the acoustic source and the receiving signals 
\cite{carter}. On similar lines, phase synchronisation is used to 
estimate the time delay between and among the atmospheric variables observed
at 
different meteorological sites \cite{diego}. In section 3, we apply this method to afore mentioned 
theoretical models. After validating the method by the results of standard 
models, in section 4, we apply this method to estimate the time delay between 
the simultaneously recorded EEG and EMG data of essential tremor subjects. We 
discuss the results and conclude in section 5. 
 
\section{Methodology} 
Let $x(t)$ and $y(t)$ be two simultaneously recorded data sets of length 
 $N$. The mean and standard deviation of the two data sets are, respectively, 
 set to zero and one. We divide the data sets into $M$ disjoint 
 segments of length $L$, such that $N=L \cdot M $. We calculate power 
 spectra, $\widehat{S_{xx}}$, $\widehat{S_{yy}}$ and cross spectrum 
 $\widehat{S_{xy}}$, 
  which is the  
 Fourier transform of the cross-correlation function of the signals  $x(t)$ and
 $y(t)$ \cite{ml1}, in each segment. Over cap in all the 
 quantities indicates that it is an estimate of that quantity. Finally, we 
 average  
 the power spectra  
 and the cross-spectrum across all the segments and calculate coherence as 
 follows \cite{hal1},  
$$ \widehat{C(\omega)}~=~\frac{\widehat{|S_{xy}(\omega)}|^2}{\widehat{S_{xx}(\omega)} \cdot \widehat{S_{yy}(\omega)}}~. 
\eqno 1 $$  
 The coherence spectrum provides the strength of correlation between the two 
signals, $x(t)$ and $y(t)$.  The confidence limit for coherence at the $100 
\% ~\alpha $ is given by  
$1-(1-\alpha)^{\frac1{(M-1)}}$. Thus, the coherence spectrum is always considered 
with this  
line. In all of our analysis, $\alpha$ is set to 0.99, and hence the confidence 
limit is $1-0.01^{\frac1{(M-1)}}$. The estimated value of coherence, at a 
frequency, above this 
line indicates (a) significant coherence between the two time series at this 
frequency;  
(b) the magnitude (deviation of coherence from this line) determines the
degree of linear correlation between the two 
time series at this frequency. The estimated value of coherence at a frequency below this line 
is considered as the lack of correlation between the two time series at this 
frequency.  
 
If the sampling frequency of the signals is $sf-$ $Hz$ (i.e. $sf$ number of 
data points are sampled per second), then the frequency resolution of the quantities in 
eq. 1, is $\frac{sf}{L}$.  Thus, one should optimally choose the value of $L$ 
depending on the purpose of analysis, to compromise between the sensitivity 
and reliability. Usage of the fixed segment length $L$ is questioned 
and a variable segment length is suggested in \cite{jeti1}, to make 
quantitative assessment of the signal. But, for all practical purposes, a fixed 
segment length is easily implementable and hence, it is used in the 
forthcoming analysis.

While the coherence spectrum provides the strength of correlation between the two 
signals $x(t)$ and $y(t)$, time (delay) information between the two signals can 
be obtained from the phase spectrum, which is the argument of the 
cross-spectrum \cite{ml1,hal1}, 
$$ \widehat{\Phi(\omega)}~=~arg\{\widehat{S_{xy}(\omega)\}}.  \eqno 2 $$ 
Following \cite{ml1}, eq. 2 can be further simplified to see the explicit 
appearance of the time delay in it as follows, 
$$\widehat{\Phi(\omega)}~=~\omega \delta, \eqno 3$$ 
where $\delta$ is the time delay. 
 
The phase estimate $\widehat{\Phi(\omega)}$ and its upper and lower 95\%
confidence interval are given by 
\cite{hal1} 
$$\widehat{\Phi(\omega)}~\pm 1.96~\left[\frac1{2M}\left(\frac1{\widehat{C(\omega)}}-1\right) 
\right]^\frac12. 
\eqno 4 $$  
Thus, the confidence interval of the phase estimate is inversely related to coherence.     
In the rest of the paper we use estimates of the coherence and phase without 
the over hat for the sake of convenience. 
 
One of the conventional ways to estimate a time delay, in frequency domain,  is to fit a straight 
line to the phase spectrum (eq. 3) in the frequency band of significant 
coherence, as the phase can be reliably estimated only in the frequency 
band of significant coherence (see eq. 4). This method of estimating delay 
is possible when we have a broad band coherence and limits its applicability
to the narrow band coherent signals. In some cases, coherence extends to first 
harmonic and hence the phase values at the harmonics are used to increase the 
reliability of the delay estimate \cite{pb1}.  
 
Further, if two time series show significant coherence over a wide 
range of frequency band, but have a minimal phase relation, then the 
estimation of time delay from the phase estimate is not straight 
forward. Under such conditions, phase estimate, in addition to 
the $\omega \delta $ term (see eq. 3), will also contain a  
frequency dependent factor, namely,  the argument of the transfer function, 
${\rm{arg}}~  
A(\omega)$. In such cases, ${\rm{arg}}~A(\omega)$ is estimated by Hilbert 
transform, and then delay is estimated from the phase estimate after 
subtracting the estimated transfer function from the phase estimate \cite{ml2}. 
From the above discussion it is clear that phase estimates cannot be used to 
estimate the 
delay from narrow band coherent signals which are often observed in human 
physiological data \cite{hel1,jan1}. For this purpose, we 
use the method of maximising coherence.  
 
As discussed in the last part of the introduction, a delay $\delta$
between two 
time series, will cause a reduction in the estimated coherence between
them.  In order to compensate for the reduction in coherence due to delay and
thereby to estimate the delay, we realign the time series by artificially
shifting them. For this, we shift one of the time series by a lag $(\tau)$
keeping the other constant and consider coherence at the selected 
frequency band $\omega_0$ as a function of $\tau$, $C(\tau)_{\omega_0}$ and repeat the same for the other time series. If there is a 
delay between the two time series, coherence in the selected frequency band 
$C(\tau)_{\omega_0}$ will increase from its value at $\tau=0$ and reach a maximum at the $\tau$  corresponding to 
$\delta$. After shifting one of the time series by a lag $\tau$, the length of
the shifted time series will be 
less than the un-shifted   
time series by $N-\tau$ ($\tau$ in sample units) data points. We discard the extra data points in 
the un-shifted time series to have the same length for both time series. Further, 
coherence is a relative measure and changes with the length of the data. 
It is reflected in the 99\% confidence level (see above). For each shift, 
we discard the points corresponding to the maximum lag from the entire length of the 
time series, and consider only the length of the data which is integer 
multiple of $L$. By doing so, coherence $C(\tau)_{\omega_0}$ calculated for 
all the lags will have  
the same confidence level, thereby confirming that the maximum 
value of coherence is reached only because of the delay and not due to the 
spurious effects caused in estimating coherence for different lengths 
of data. As we have confidence level as a check for the 
reliability of  
$\delta$, we don't need any additional tests like, surrogate tests 
\cite{kantz} to assess the reliability of the 
results. However, in order to get the variability in the estimated delay, we 
use surrogate analysis. 
 
Surrogate analysis is introduced in the context of nonlinear time series 
analysis to check whether or not the time series under consideration has got 
a nonlinear structure \cite{kantz,theiler}.  Making the (null) hypothesis that 
the time series has come from a linear process, several 
linear realizations of   
the time series namely, surrogates, are synthesised. Then, the original time 
series and the surrogates are quantified by a suitable 
discriminating statistic. Any deviation in the discriminating statistic 
calculated for the  
original time series and surrogates indicates the 
presence of nonlinear structure in the original time series and thereby 
rejecting the null  
hypothesis \cite{theiler}. There are different ways of preparing 
the surrogates \cite{kantz,schreiber}. The 
misconception  
of surrogate analysis is discussed in \cite{jeti2}. As mentioned above, 
the objective of the methods to generate surrogates \cite{kantz,schreiber} is to 
synthesise a new data set by  
destroying the nonlinear structure present in the original data set. But our 
aim is not to show the presence of nonlinear structure in our data. For 
the purpose of calculating the error-bars of delay, surrogates are generated by exploiting one of the basic 
assumptions of spectral analysis that distinct parts of the time series are 
independent \cite{hal1}. Instead of shuffling the time series as a whole (which 
is done in one of the methods to synthesise surrogates, amplitude adjusted surrogates) we 
shuffle the disjoint data segments from which the original spectrum is 
estimated (see above). We shuffle only the un-shifted time series (see above) 
from which the information is assumed to flow to the other time series (which 
is time shifted in advance in maximising coherence analysis) with a delay. By
doing so, the spectra of both the time series, 
$\widehat{S_{xx}(\omega)}$ and $ \widehat{S_{yy}(\omega)}$ in 
eq. 1 will remain the same but the cross spectrum $\widehat{S_{xy}(\omega)}$ 
in eq. 1 will be different. This type of surrogate is similar to the one 
proposed in  \cite{schreiber} where the whole time series is shuffled but by 
preserving two point correlation (i.e. power spectrum) of the original time 
series.  For all the analyses reported in this paper we synthesise 19 
different realizations of surrogates. We make a null hypothesis that
$C(\tau)_{\omega_0}$ obtained is due to spurious correlations between the two
time series. For each realization of surrogate we calculate time 
delayed coherence $C(\tau)_{\omega_0}^{surr}$. We calculate the significance
of difference $S(\tau)$ between the $C(\tau)_{\omega_0}^{surr}$ calculated for
surrogates and $C(\tau)_{\omega_0}$ where $S(\tau)
=\frac{\left|C(\tau)_{\omega_0}-<C(\tau)_{\omega_0}^{surr}>
\right|}{\sigma[C(\tau)_{\omega_0}^{surr}]}$, where $<.>$ indicates the average 
over different realizations of surrogates and $\sigma[.]$ indicates the standard deviation 
between different realizations of surrogates. Any value of $S(\tau) > 2 $
indicates that the
$C(\tau)_{\omega_0}$ obtained is not due to spurious correlations
\cite{theiler} and hence the null hypothesis is rejected. 
If the
null hypothesis is rejected we consider $C(\tau)_{\omega_0}$ for further analysis 
otherwise we discard it as spurious correlation. For the 
$C(\tau)_{\omega_0}$ qualified in the surrogate analysis (for which null
hypothesis is rejected) we calculate the 
error in the delay in the following way: we subtract $C(\tau)_{\omega_0}^{surr}$ 
calculated for each realization from $C(\tau)_{\omega_0}$ and calculate delay 
for each subtracted function (i.e.) 
$C(\tau)_{\omega_0}-C(\tau)_{\omega_0}^{surr}$. We report here the delay of 
the system as the mean value of the delays (calculated for the 19 surrogate 
subtracted realizations) and their standard deviation as error-bar. As the
increase in the  $C(\tau)_{\omega_0}$ when compensated for the delay is very
small, for the sake of clarity we plot $C'(\tau)_{\omega_0} = \left[
C(\tau)_{\omega_0}-<C(\tau)_{\omega_0}^{surr}>\right]
-\left[C(\omega_0)-<C(\omega_0)^{surr}>\right]$. Note that $C(\omega_0)$ is
the value of $C(\tau)_{\omega_0}$ at $\tau=0$ and $<C(\omega_0)^{surr}>$ is
the average of $C(\tau)_{\omega_0}^{surr}$ for different realizations of the
surrogates at $\tau=0$. By this definition $C'(\tau)_{\omega_0}$ will pass
through a zero value at $\tau=0$ and show a maximum value (above zero) at
$\tau=\delta$ (for $\delta \ne 0$) between the  processes. Though we calculate $S(\tau)$ for different 
$\tau$ values at which $C(\tau)_{\omega_0}$ is evaluated, we report here
$S(\tau=\delta)$ as it is the most relevant one in determining the
significance of the $C'(\tau=\delta)_{\omega_0}$ (and hence $C(\tau=\delta)_{\omega_0}$). Based on the above arguments
the delay between the two time series is given by:
$$\delta = {\max_\tau}~C'(\tau)_{\omega_0}.$$

\section{Application to coupled R\"ossler attractors} 
In this section, we apply the method of maximising coherence, to  
coupled R\"ossler systems. The dynamics of the $i$-th attractor is given as 
follows: 
 
\begin{eqnarray} 
\dot{x}_i(t) &=& -[y_i(t)-z_i(t)]~+~\epsilon_{j,i}~x_j(t-\delta) \nonumber \\ 
 \dot{y}_i(t)&=&x_i(t)~+~a~y_i(t) \nonumber \\ 
 ~\dot{z}_i(t)  &=&b~+~z_i(t)~[x_i(t)~+~c] \nonumber , 
\end{eqnarray} where $a~=~0.38,~b~=~0.3,~c~=~4.5$, are the 
parameters. Coupling is established through the $x$ component between the 
attractors. $\epsilon_{j,i}$, is the coupling constant and $x_j$ is the $x-$ 
component of the $j-$th attractor. In our study, we consider the dynamics of two 
coupled R\"ossler attractors, and hence $i~=~1$ and $j~=~2$ and $\delta$ is 
the delay. Further, we don't consider the self coupling of the oscillators.  
\begin{figure}[H] 
  \begin{center} 
    \includegraphics[width=2in]{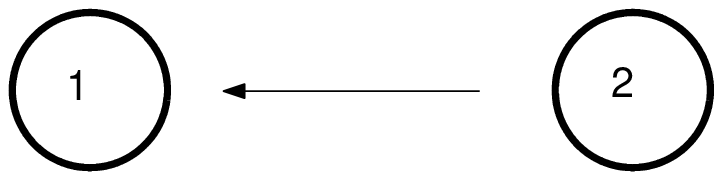} 
    \caption{Coupling scheme of the attractors 1 and 2. $\epsilon_{2,1}=0.16$ and 
      $\epsilon_{1,2}=0$. $\delta$ is set to 2 sec.} 
    \label{fig1} 
  \end{center} 
\end{figure}  
First, we consider the coupling scheme as shown in Fig. 1, where two 
R\"ossler attractors are coupled and the information flow is from 
attractor 2 to attractor 1 ($\epsilon_{2,1}=0.16$ and $\epsilon_{1,2}=0$) as 
governed by the above set of equations. $\delta$ is set to 2  
sec. The above equations are simulated by Euler scheme with a step size of 
0.01 sec and for the subsequent analysis the data are down sampled to 0.1 
sec. As the coupling is established through the $x-$ component between the 
attractors, we use the $x-$ time series to perform the delay estimation. We 
have used 30000 data points for further analysis. 
   
Power spectra of the two attractors are shown in Fig 2a. Solid line 
represents the power spectrum of attractor 1 and solid line with 
dots represents the power spectrum of attractor 2. For the sake of 
clarity the spectrum of attractor 1 is shifted vertically upwards by a factor
of 5 from 
its original position. We have used a segment length 
of $L=1000$ and hence the frequency resolution (of the quantities in eq. 
1) is 0.01 $Hz$. Spectra show dominant activities in the frequency range 
of 0.1 to 0.2 $Hz$, which 
can be related to the mean orbital period of $\sim5$ sec of the 
attractor \cite{nar1,wolf}. Since the dynamics of the attractor 1 is 
perturbed by coupling (see Fig. 1), its spectrum looks slightly 
different from that of the attractor 2. The coherence spectrum of the 
coupled attractors is shown in Fig. 2b. Horizontal line around the coherence of 0.15 indicates 
the confidence level of $\alpha=0.99$.

In the frequency band between 0.1-0.2 $Hz$ (see Fig. 2a), where the 
individual attractors show dominant activities, the coupled systems show 
significant coherence. Estimated phase (eq. 2) between the coupled 
attractors is shown in Fig. 2c.  Error-bars represent the 95\% confidence
interval 
of the phase estimate given by eq. 4. Thus the phase can be reliably 
estimated in the frequency band (0.1-0.2 $Hz$), where the coupled 
attractors show a significant coherence (see Fig. 2b). The result of
maximising coherence is shown in Fig. 2d where $C'(\tau)_{\omega_0}$ is
plotted as a function of lag $\tau$. We have chosen $\omega_0 = 0.13 Hz$
(and is shown in Fig. 2b) from the basic frequency of the oscillators. Solid line at
$C'(\tau)_{\omega_0}=0$ indicates $C(\omega_0)-<C(\omega_0)^{surr}> $(see the
last part of the methodology for details).
Negative 
shift indicates that the time series of attractor 1 is in advance and positive 
shift indicates that the time series of attractor 2 is in advance. 
Since there is a delay of 2 sec, the coherence calculated at $\tau=0$ is 
zero (see above). In order to compensate for the reduction in the coherence and
thereby to estimate the time delay, time series of the attractors are shifted
by a time lag  $\tau$. As the delayed information flows from attractor 2 to
attractor 1, coherence  $C'(\tau)_{\omega_0}$ increases from zero (see above) when the time series of attractor 1 is
shifted \textbf{back} in time and reaches a maximum at the $\tau=-2.1$ which is the
time delay between the two time series ($\delta$ used in the simulation is 2
sec). The significance of deviation from the surrogate data $S(-2.1)$ is
2.24 and indicates that
$C'(\tau=-2.1)$ is not due to spurious correlations.
The error-bar for the delay (as 
explained in the methodology section) is calculated as 0.4 sec. Thus, the 
time delay $ \delta $ is $-2.1 \pm 0.4 $ sec and is well captured by the 
maximising coherence analysis (expected delay value is 2 sec for the flow from
attractor 2 to attractor 1). 
\begin{figure}[H] 
  \begin{center} 
    \includegraphics[width=5in,height=4.5in]{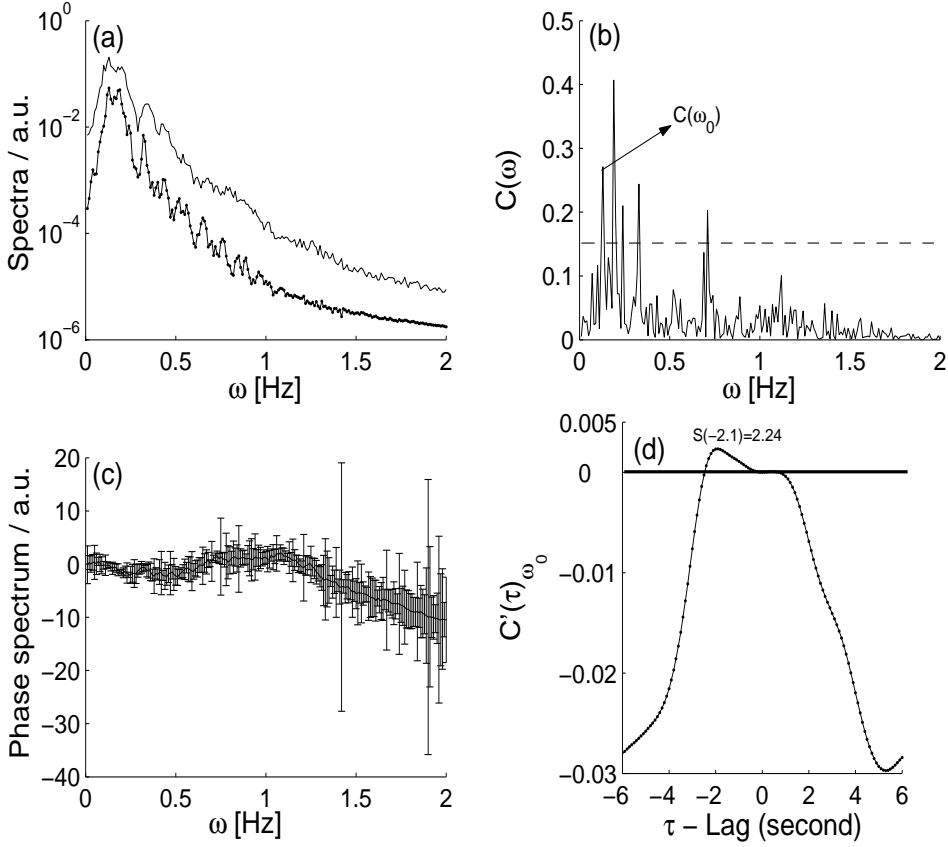} 
    \caption{(a). Solid line represents power spectrum of attractor 1 and
    solid line with dots represents power spectrum of attractor 2. For the sake
    of clarity the power spectrum of  
      attractor 1 is shifted vertically upwards by a factor of 5 from its original
    position. (b). Estimated 
      coherence spectrum of the coupled R\"ossler attractors with 
      $\delta=2 $ sec.  Horizonal line at the coherence of 0.15 indicates the confidence level 
      ($\alpha=0.99$).  There is a significant coherence in the 
      frequency range of 0.1-0.2 $Hz$ between the two attractors. The value of
    $\omega_0$ used for maximising coherence analysis is indicated along with
    its 
    coherence value $C(\omega_0)$. (c). Phase estimate 
      $\Phi(\omega)$ of the coupled attractors.  Error-bars represent 
      95\% confidence interval defined by eq. 4. Errors are relatively small in 
      the region of significant coherence which is in accordance to eq. 
      4. (d). Coherence $C'(\tau)_{\omega_0}$, as the function of lag 
      $\tau$. Horizontal line at  $C'(\tau)_{\omega_0}=0$ indicates
    $C(\omega_0)-<C(\omega_0)^{surr}>$. $<.>$ represents the average over
    different surrogates. $C'(\tau)_{\omega_0}$ has reached the maximum value at the $\tau$ (delay) value of -2.1 sec 
      (from attractor 2 to 1), and the expected value of $\delta$ is 2 sec for
    the flow from attractor 2 to attractor 1. The significance of deviation
  from the surrogate $S(-2.1)$ is 2.24 and indicates that $C'(\tau)_{\omega_0}$ is not due to spurious
    correlations.}
 \label{fig2} 
  \end{center} 
\end{figure} 
Next, we consider the coupling scheme as shown in Fig. 3, where two R\"ossler 
attractors are coupled as considered above, but with a bi-directional flow, 
$\epsilon_{2,1}=0.15$ and $\epsilon_{1,2}=0.1$. $\delta$ used in the coupling 
is same as in Fig. 1, which is 2 sec.  
 \begin{figure}[H] 
  \begin{center} 
    \includegraphics[width=2in]{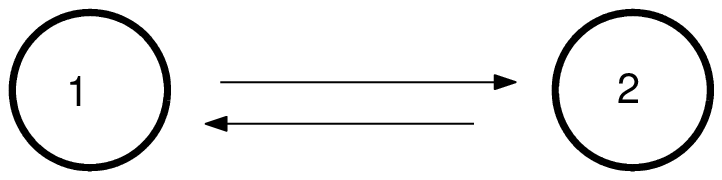} 
    \vskip 0in 
    \caption{Coupling scheme of the attractors 1 and 2. $\epsilon_{2,1}=0.15$ and 
      $\epsilon_{1,2}=0.1$. $\delta$ is set to 2 sec.} 
    \label{fig3} 
  \end{center} 
\end{figure} 

Power spectra of the attractors look 
almost the same as the two attractors perturb each other almost to the same 
extent (see Fig. 4a).  
As obtained for uni-directional coupling, in this case 
also, significant coherence is between 0.1-0.2 $Hz$ (Fig. 4b). As the number
of data points used is the same (30000) as in the uni-directional coupling,
the 
confidence level $\alpha=0.99$ is also around the coherence of 0.15 (see the 
horizontal line in Fig. 4 b which is same as in Fig. 2b). Phase estimate of 
the bi-directionally coupled systems is shown  
in Fig. 4c with the 95 \% confidence interval as the error-bars. Error-bars are  
relatively narrow (see eq. 4) in the frequency band of 0.1-0.2 $Hz$ where 
significant coherence is observed (as seen in Fig. 2c). Result of the
maximising coherence for bi-directionally coupled systems is shown in
Fig. 4d.  The horizontal line at $C'(\tau)_{\omega_0}=0 $  has the meaning as
in Fig. 2d. In this case 
0.15 $Hz$ is used as $\omega_0$ (and is shown in Fig. 4b) which is the basic
frequency of the attractors.  $C'(\tau)_{\omega_0}$ shows two maxima, one at $\tau=-2.5$ sec which 
corresponds to the flow from attractor 2 to 1 and another at $\tau=  1.7$ sec 
which corresponds to the flow from attractor 1 to 2, while the $\delta$ used 
in the coupling is 2 sec in both directions (see Fig. 4d). Also, in this case, 
the significance of deviation from surrogate calculated by $S(\tau)$ at $\tau$
values -2.5 sec and 1.7 sec are well above 2 (see Fig. 4d) indicating that the
$C'(\tau)_{\omega_0}$ obtained at these two $\tau$ values are not due to
spurious correlations.  
Thus, the time delays of 
the system are $-2.5 \pm 0.5 $ sec (for the flow from attractor 2 to 
attractor 1)  
and $1.7 \pm 0.4 $ sec (for the flow from attractor 1 to attractor 2) and are 
well within the expected value of 2 sec in either direction (see Fig. 3). The
error-bars are obtained by surrogate analysis.
\begin{figure}[H] 
  \begin{center} 
    \includegraphics[width=5in,height=4.5in,angle=0]{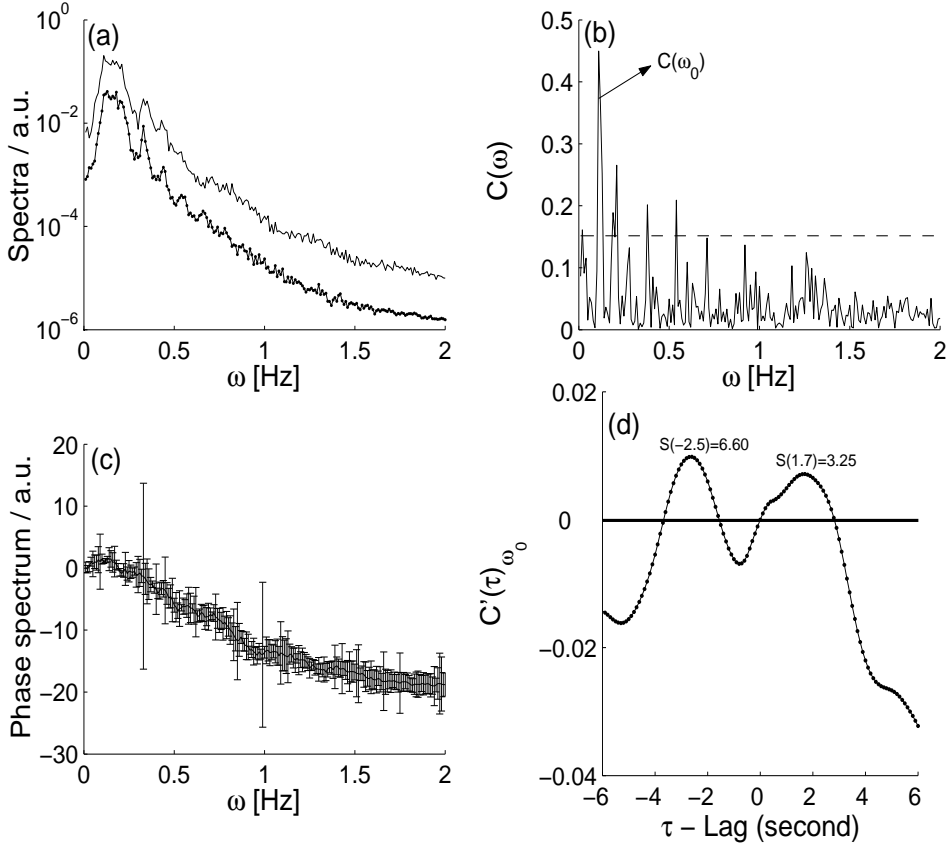} 
    \caption{(a-c). Explanations are as in Fig. 2 but for bi-directionally coupled 
      R\"ossler attractors, $\epsilon_{2,1}=0.15$ and 
      $\epsilon_{1,2}=0.1$. (d). Result of maximising coherence
    $C'(\tau)_{\omega_0}$.
    $C'(\tau)_{\omega_0}$ has reached 
      maximum value at two places, one at -2.5 sec (from attractor 2 to 
      1) and another at 1.7 sec (from attractor 1 to 2), where 
      $\delta$ is set to  2 sec in both directions. The significance of
    deviation from surrogate at $S(-2.5)$ is 6.60 and at $S(1.7)$ is
    3.25 and indicates that $C'(\tau)$ at $\tau$ values -2.5 sec and 1.7 sec
    are not due to
    spurious correlations. The horizontal line at  $C'(\tau)_{\omega_0}=0 $
    has the same meaning as in Fig. 2d.} 
    \label{fig4} 
  \end{center} 
\end{figure}
Results of the theoretically coupled systems are summarised in Fig. 5. For the 
unidirectional flow, shown in Fig. 1, which is referred to as scenario 1 in 
Fig. 5, the value of the delay obtained by maximising coherence, shown by '$ 
\ast$' with the error-bars (calculated by surrogate analysis) 
shown as vertical line, is close to the expected value of -2 sec (shown as dot in 
Fig. 5). Similarly for the bi-directional flow (see Fig. 3) which is referred 
to as scenario 2 in Fig. 5, the values of the delay obtained by maximising 
coherence with their error-bars (vertical lines) are close to the expected
delay 
of 2 sec in either direction (see Fig. 3). Compared to the unidirectional 
flow, the results of the bi-directionally coupled oscillators show slightly 
more deviations from the expected value of 2 sec but when we include the 
error-bar, they are close to the expected value. So when dealing with the 
biological systems we should consider the delay always with the error-bar 
before we draw any conclusion from it. For the uni-directionally coupled
oscillators there is no significant change in the result when $\omega_0$ is chosen as
0.19 $Hz$ where there is a significant coherence between the two oscillators. Further for the above studied systems
similar results can be obtained by  
cross-correlation analysis and by the time delayed phase synchronisation analysis
as well \cite{diego}.  
\begin{figure}[H] 
  \begin{center} 
    \includegraphics[width=4in,height=3in,angle=0]{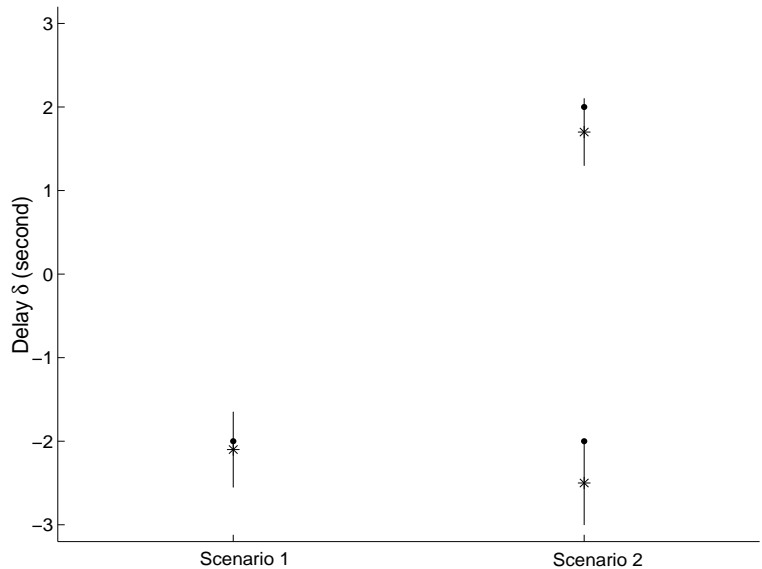} 
    \caption{Results of delay estimation by maximising coherence for 
    theoretically coupled systems. Negative delay indicates the flow from
    attractor 2 to attractor 1 and the positive delay indicates the flow from
    attractor 1 to attractor 2. Scenario 1 represents uni-directionally 
    coupled oscillators (flow from attractor 2 to attractor 1). Scenario 2 
    represents bi-directionally coupled oscillators. Expected delay values 
    are shown by dots. Delay values obtained by maximising coherence are shown 
    by '$\ast$'. Vertical lines are the error-bars of delay (see text for 
    details).} 
    \label{fig5} 
  \end{center} 
\end{figure} 
 
\section{Application to EEG and EMG data} 
Tremor is an involuntary, periodic movement of the parts of the body 
\cite{deuschl}. It can be classified as normal and pathological tremor 
depending on amplitude and the conditions under which tremor is activated 
\cite{deuschl}. Essential tremor is a common movement disorder characterised 
mostly by postural tremor of the arms. Other neurological abnormalities are
typically 
absent in essential tremor \cite{findley}. It has been shown by experimental
studies in animals 
\cite{llinas,lamarre} and by Positron emission tomography 
or functional magnetic resonance imaging studies 
\cite{hallet,jenkins,bucher} on human 
beings, that different parts of the brain are involved in essential tremor.  
The correlation between the thalamic activity and forearm electromyogram 
provides a direct evidence for the involvement of the thalamus in the  
tremor oscillations \cite{hua}. Further, control of essential tremor by 
stereo-tactic lesions and high frequency stimulation of the ventrolateral 
thalamus adds support for the involvement of thalamus in the 
mechanism of essential tremor \cite{benabid,schuurman}. As the cerebellum has
its main outputs to thalamus which in turn has strong projections to
cortex, it is hypothesised that essential tremor arise from an oscillating
cerebello-thalamo-cortical loop the output station of which is the motor
cortex \cite{hel1}.
\begin{figure}[H] 
  \begin{center} 
    \includegraphics[width=5in,height=4.5in,angle=0]{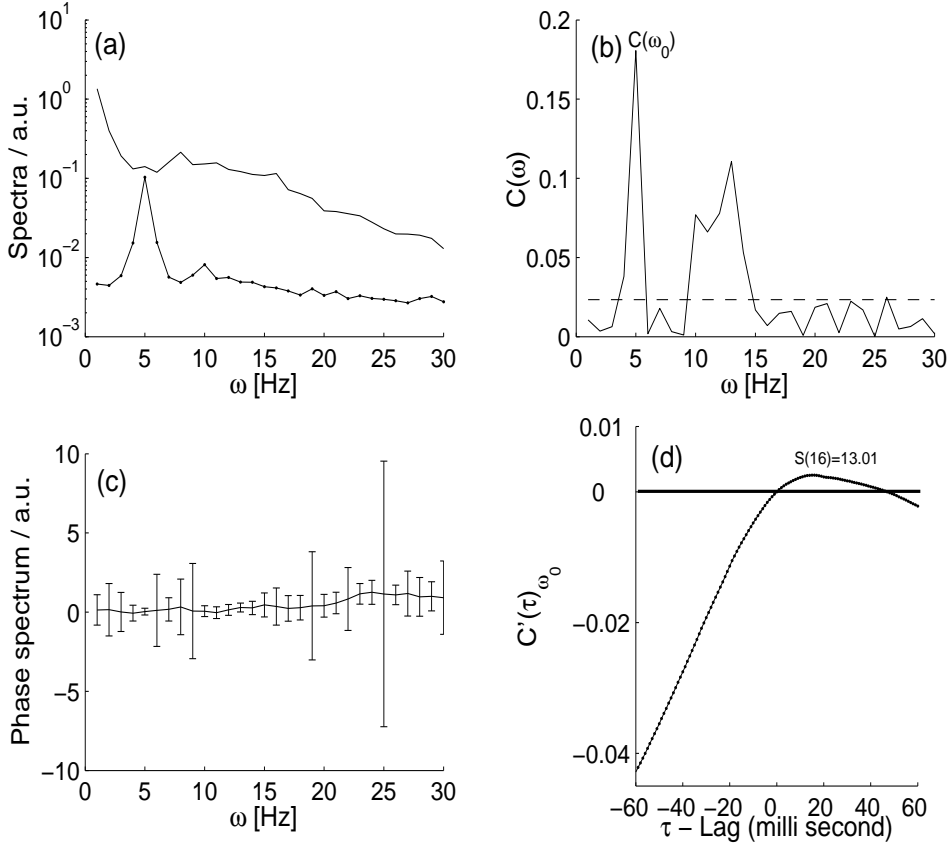} 
    \caption{Results of spectral analysis and delay estimation for 
      subject 1. (a). Solid line represents power spectrum of C4 of EEG
    (the central electrode overlaying the primary motor cortex, PMC) and solid line with dots 
    represents the power spectrum of extensor muscle 
      with a frequency resolution of 1 $Hz$. For the sake of clarity, the 
    power spectrum of EEG is shifted vertically upwards from its original 
    position by a factor of 10. (b). Coherence spectrum of C4 
      and extensor muscle. The horizontal line at the coherence of 0.024 
      indicates the confidence level of $\alpha=0.99$. There is a 
      significant coherence at the (tremor) frequency of 5 
      $Hz$ and at 10-15 $Hz$ as well. The value of $\omega_0$ used for the
    maximising coherence is shown along with its coherence value $C(\omega_0)$. (c). Phase estimate of the C4 and 
    extensor muscle with 95\% confidence interval as the error-bars. Error-bars are narrow at the 
      significantly coherent frequency bands of 5 $Hz$ and 10-15 $Hz$ as well. (d). Coherence as a 
      function of $\tau$, $C'(\tau)_{\omega_0}$. It has reached maximum 
      at $\tau=16$ $ms$. The horizontal line at $C'(\tau)_{\omega_0}=0 $ has
    the same meaning as in Fig. 2d. The significance of deviation
    from the surrogate is given as $S(16) = 13.01$.} 
    \label{fig6} 
  \end{center} 
\end{figure}  
Cortico-muscular coherence for essential tremor has been studied earlier \cite{halliday} but has
not been ascertained at the tremor frequency in
\cite{halliday}.  We show below that there is a significant 
cortico-muscular coherence when the signal to noise ratio of the tremor 
increases beyond a certain threshold value as seen in 
\cite{hel1,hellwig2}.  
Since the paper is aimed at time delay estimation, the detailed 
results and discussion of the spectral analysis of the tremor study will be 
reported elsewhere \cite{jan1}. 
 
For this purpose we consider five patients (with mean age 66 years and
standard deviation 5.1 years), which showed postural essential tremor of the
arms. In a 
dimly lighted room, patients are asked to sit on a comfortable chair in a 
slightly supine position with both hands held against gravity while the
forearms are supported. EEG is 
recorded with a 64-channel EEG system (Neuroscan) with standard electrode
positions  \cite{klem}. Surface EMG is recorded 
from EMG electrodes attached to wrist flexors and extensors of both 
arms.  For all the subjects EEG and EMG are sampled 
at a rate of 1000  
$Hz$. EEG and EMG are bandpass filtered online, respectively, between 
0.01-200 $Hz$ and 30-200 $Hz$. Data are stored in a computer and are analysed offline. In all the cases EMG is recorded on both 
hands. Each recording lasted for 1-4 minutes. Artifacts like base line shift, 
eye blinks etc. are discarded by visual inspection. 
 
Before further analysis, EMG signals are full wave rectified (magnitude of 
the deviations from mean) 
and EEG is made reference free by constructing the second (spatial) derivative, 
(Laplacian) \cite{hj1,hj2}.  We discuss in detail for one of the subjects, 
subject 1,  
which we feel typical and for the remaining subjects we summarise the results in a 
table.  
 
For subject 1, the power spectra of EEG (solid line) and EMG (solid line with dots), see 
Fig. 6a, are 
calculated with $L=1000$ and hence the frequency resolution is 1 $Hz$ (since 
$sf=1000 $ $Hz$). In Fig. 6a, we show the power spectra of one of  
the EEG channels (C4) in 
the contralateral (right) side to tremor (left) hand and the rectified EMG 
(extensor muscle) of the same hand.  Power spectrum of EEG (shown as solid 
line in Fig. 6a) is shifted vertically upwards by a factor of 10 from its initial 
position for the sake of 
clarity. Power spectrum of EEG shows dominant 
activity in a broad band between 8-20 $Hz$ containing the bands of 10-13 $Hz$  
and  15-20 $Hz$, which correspond, respectively, to the alpha and beta activity 
of the brain \cite{guy}.  Muscle spectrum (solid line with dots in Fig. 6a) shows tremor activity around 5 
$Hz$. Correlation between the muscle and C4 (the central electrode
overlaying the primary motor cortex, PMC) of EEG cannot be guessed 
from the EEG spectrum as there is no significant activity around 5 $Hz$ (tremor
frequency).  Correlation becomes clear when we consider coherence. Figure 6b 
shows coherence between C4 and extensor muscle. The horizontal line around the 
coherence of 0.024 indicates the confidence level of $\alpha=0.99$. There is a 
significant coherence (and hence correlation) between C4 and the tremor 
activity reflected by the muscle \cite{hel1,hellwig2,mima1,mima2,mima3,hal3} 
and also in the frequency band of 10-15 $Hz$. 
Phase estimate of this system is given in Fig. 6c with $95 \%$ confidence interval as 
error-bars. It is clear from Fig. 6c that phase can be reliably estimated at 
the (tremor and the significant coherence) frequency of 5 $Hz$ (confidence
interval is narrow) and also in the frequency band of 10-15 $Hz$. In
an earlier study \emph{Hellwig et al.} \cite{hellwig2} investigated time
delays based on the value of the phase estimate $\widehat{\Phi(\omega)}$ at
the \textbf{tremor frequency}. But the results obtained \cite{hellwig2} are
not in the interpretable level.  The reason postulated in \cite{hellwig2} that
the delay might be modulated by frequency dependent mechanism is hardly
conceivable in a narrow (single) frequency band at which phase estimate is
significant. The reason may be due to the weak assumption of the model
$\widehat{\Phi(\omega)} = \omega\delta + \theta$, where $\theta$ is the
indispensable constant \cite{mima1,mima2}, phase shift between the two
processes which is ignored in
\cite{hellwig2}. However, we show below that the delay estimated by maximising
coherence is reasonably in good agreement with the experimentally observed
conduction velocity and is in physiologically interpretable range.

Application of the maximising coherence method to real life data like EEG is 
not   
straight forward. Since the maximising coherence method is aimed at to capture
the time delay (by making up for the reduction in the coherence) by
artificially shifting (local dynamics) the time series, it is very sensitive
to non-stationarities in the time series. However, in order to make coherence
analysis robust against non-stationarities within the time series, we use the
following way to discard the non-stationary parts of the data:    
We calculate autocorrelation function 
\cite{kaplan}, in each of the segments (of length 1000 samples). Then the 
de-correlation time, time at which autocorrelation function falls to $exp(-1)$ 
(as the first value of autocorrelation function is 1 by definition) 
 \cite{kaplan} is calculated for all the segments. Non-stationarities due to 
base line shift, eye blinks will introduce artificial trends (lifting up of the 
mean value of the signal) \cite{chen} in the signal, which will have higher de-correlation 
time than the stationary parts of the signal. In order to discard the 
non-stationary parts, about 50 \% of the median of 
the de-correlation time calculated in all the segments is taken as threshold 
and parts of the EEG and EMG data with de-correlation time greater 
than the above defined threshold are discarded in the delay estimation. After 
discarding  
the non-stationary parts of the data, the segments left out on either side of 
the removed parts are stitched as if they are recorded continously. Of course, 
care should be taken in applying the above method. The above method will 
capture only the non-stationary parts if it is applied to the data of low 
signal to noise ratio which can be avoided by visual inspection of the data.  

Result of the delay estimation between C4 and muscle, is shown in Fig. 
6d. The horizontal line at $C'(\tau)_{\omega_0}=0$ in Fig. 6d has the same
meaning as in Fig. 2d. 
Here, we have used $\omega_0=5$ $Hz$ (see Fig. 6b), where 
significant coherence is observed. Negative shift indicates that the time series 
of EEG is in advance. In Fig. 6d, $C'(\tau)_{\omega_0}$ reaches a 
maximum value at $\tau=16$ $ms$, indicating the cortex (time series of EMG 
is in advance for positive values of $\tau$) is driving the 
tremor. The significance of deviation from the surrogate is given by
$S(16)= 13.01$ and indicates that $C'(\tau)_{\omega_0}$ obtained is not
due to spurious correlations.
The error-bar obtained by surrogate analysis is 7 $ms$ and hence the delay
obtained
for C4 is 16$\pm$7 $ms$ which is in good agreement with the experimentally 
observed value of around 15 $ms$ \cite{roth}. Comparison of Fig. 6d with
Fig. 2d, 
shows that the information flow is in one direction (i.e.) from 
cortex to muscle. But in other subjects, in (an electrode from)
this region, we have also observed a bi-directional flow, a situation
similar to Fig. 4d.
\begin{figure}[H]
  \begin{center}
    \includegraphics[width=5.5in]{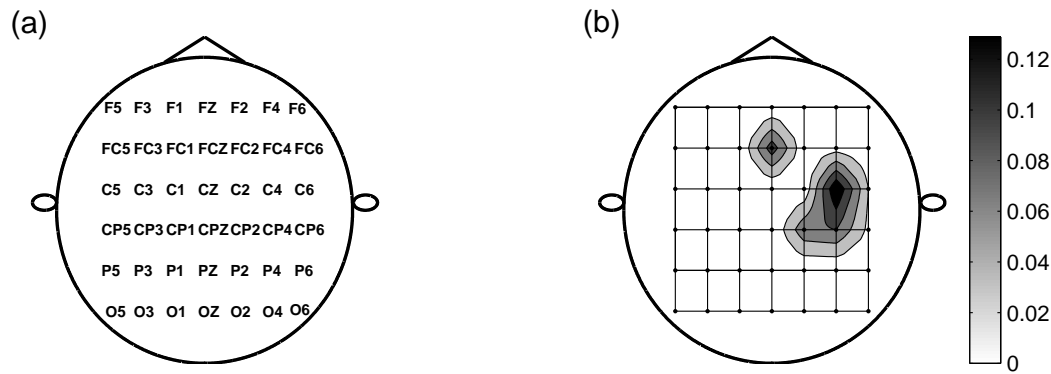}
    \caption{(a). A part of the 64-channel arrangement on the surface
      of the scalp used to record EEG. Electrodes on the left hand side
      are odd numbered and those on the right hand side are even
      numbered. (b).  Isocoherence map of the subject 1, constructed by
      considering the coherence at the tremor frequency (5 $Hz$ in this
      case). Magnitude of the coherence is indicated by the gray scale-bar
      juxtaposed to the isocoherence map. Parts of the scalp which
      showed significant coherence are coloured black and the
      confidence level of $\alpha=0.99$ is marked with white colour (which is
      set to zero, see text for details). (Electrodes O5-O6 are just
      displayed and are not used for coherence analysis).}
    \label{fig7}
  \end{center}
\end{figure}
Of the 64-channels of EEG recorded, coherence and delay estimation are
explained in detail for C4, as it is overlaying the primary
sensorimotor cortex which is involved in motor control of the contralateral
hand affected by essential tremor. For the rest of the electrodes (see Fig.
7a) we give the results of the coherence analysis as isocoherence
(region of scalp with same coherence) map.  Electrodes on the boundary
of the scalp are not considered for the analysis as they had low signal
to noise ratio.  Further, we have not used the electrodes
O5 - O6 for the study.  To construct isocoherence map, we have considered
the coherence $C(\omega)$ at the (tremor) frequency of 5 $Hz$ for all the
thirty five channels (see above) shown in Fig. 7a, from F5 (top left of
scalp, Fig. 7a), to P6 (bottom right of scalp, Fig. 7a). We have
subtracted the confidence level ($\alpha=0.99$) from the coherence so
that the confidence level for coherence is zero. In Fig. 7b,
surface of the scalp exhibiting significant coherence is marked with black
colour and the coherence of the parts which do not cross the confidence
level of $\alpha=0.99$ (which are set to zero after the above
normalisation) are marked with white colour.  In Fig. 7b, C4 shows maximum
coherence,  (see
the gray scale-bar juxtaposed to Fig. 7b). Thus, there is a hot spot
(region of significant coherence) in the PMC region. Similar results have
been obtained in the spectral study of essential tremor
\cite{hel1,hellwig2}.  

There is another hot spot for subject 1 in the frontal region of the
scalp (FCZ). The detailed discussion of this will be dealt with in
\cite{jan1}.
\begin{figure}[H]
  \begin{center}
    \includegraphics[width=4in,height=5.5in,angle=270]{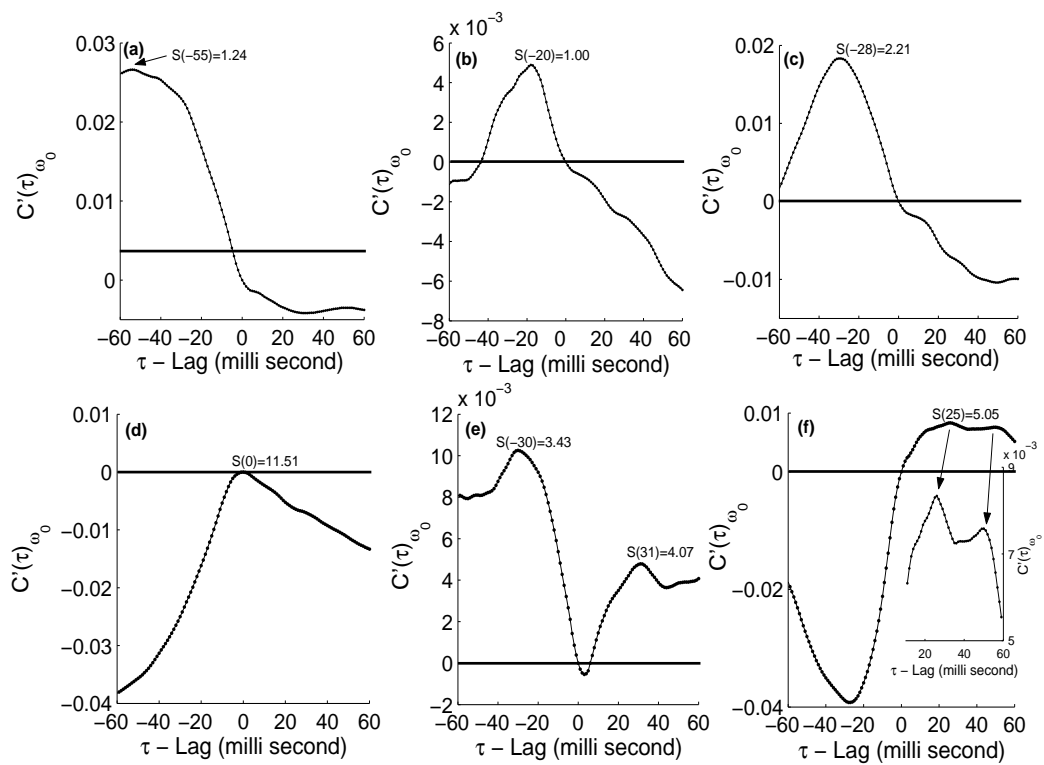}
    \caption{Results of delay estimation by maximising coherence for (a). C2
    (b). C6 (c). CP2 (d). CP4 (e). CP6 of the central hot spot and (f). FCZ of the
    frontal hot spot. The horizontal line at $C'(\tau)_{\omega_0}=0$ in Fig. 8
   (a-f) has the same meaning as in Fig. 2d. The
    significance of deviation from the surrogate $S(\tau)$ 
at $\tau=\delta$ is given in all the plots. 
 Estimated values of delay by maximising
    coherence are (c) -28 $ms$ (d) -2 $\sim$ 0 $ms$ (e) -30 $ms$ and 31 $ms$ (f)
    25 $ms$. As the maximum of $C'(\tau)_{\omega_0}=25~ms$ in Fig. 8f is not clearly
 discernable (see text for details) it is magnified in the inset shown in
 Fig. 8f. Negative shift indicates that the time series of EEG is in
    advance and positive shift indicates that the time series of EMG is in
    advance. A negative delay indicates the flow from muscle to EEG and a
    positive delay indicates the flow from EEG to muscle. C2 and C6 do not
    qualify in the surrogate analysis as their $S(\tau=\delta)$ values lie
 below 2 
 indicating that the values of $C'(\tau)_{\omega_0}$ (see Fig. 8a and 8b) obtained for these
 two electrodes are due to
 spurious correlations and therefore are not considered for further analysis. }
    \label{fig8}
  \end{center}
\end{figure}
Spectral and delay estimations for C4 and extensor
muscle have been analysed in detail showing a uni-directional
flow from PMC to muscle, with a delay of 16 $ms$. For the rest of the 
 electrodes (C2, C6, CP2, CP4, CP6) of the
central hot spot and FCZ of the frontal hot spot, the results of delay
estimation are given in Fig. 8. The significance of deviation from surrogate
$S(\tau)$ at $\tau=\delta$ is given in all the plots, see Fig. 8(a-f). For C2
(Fig. 8a) and C6 (Fig. 8b) the value of $S(\tau=\delta)$ is below two
indicating that the values of $C'(\tau)_{\omega_0}$ obtained for these two
electrodes are 
due to spurious correlations. 
Therefore we do not consider them for further analysis. For CP2,
CP4, CP6 and FCZ, $S(\tau=\delta)$ values obtained are above 2 (see Fig. 8(c-f))
indicating that  $C'(\tau)_{\omega_0}$ values obtained for these electrodes are not due to spurious correlations.
Delay values exhibited by
$C(\tau)_{\omega_0}$ shown in Fig. 8(c-f) are CP2 -28 $\pm$ 3.3 $ms$ (see
Fig. 8c), CP4, -2  $\sim$ 0 $ms$ (see Fig. 8d), CP6 -30 $\pm$ 7.4 and 31  $\pm$ 13.5 $ms$ (see
Fig. 8e) and FCZ 25  $\pm$ 10.4 $ms$ (see Fig. 8f). The maximum in
$C'(25)_{\omega_0}$ for FCZ shown in Fig. 8f is not clearly discernable and
hence this region is magnified in the inset shown in Fig. 8f where the maximum at
$\tau=25~ms$ can be seen without any ambiguity. The large
fluctuations in $C'(\tau)_{\omega_0}$ for FCZ (Fig. 8f) and C4 (Fig. 6d)
indicate larger uncertainty in the delay estimated by this method and is
one of the shortcomings 
of this method. However, this uncertainty, to some extent is revealed by the
error-bars (see above) obtained by the surrogate analysis. A positive value of
$\tau$ 
indicates that the time series of muscle is in advance and a negative
value of  $\tau$ indicates that the time series of EEG is in
advance. Error-bars are obtained by surrogate analysis as discussed in the
methodology section. A zero delay for CP4 (see Fig. 8d) may be due to the
nullification of two strong counter acting forces (drive from PMC to muscle
and opposing drive from muscle to PMC, a situation similar to phase
locking in coupled oscillators). The positive values of estimated delays including the error-bars are well in agreement with the
experimentally observed value of 15 $ms$ for forearm muscles \cite{roth}. For
the negative delay (flow from muscle to EEG), it is at least partly in keeping
with normal median Somatosensory Evoked Potential (SEP) latencies of around 20 $ms$. 

Further, in order to make the delay statistically significant we report the
delay of the hot spot as a whole as the delay of the system, by weighting the delay obtained for each
electrode (when it is qualified in surrogate analysis) by a fraction of the coherence
$C(\omega_0)$ value that each electrode contributes to the hot spot. By doing
so, electrodes showing higher tremor related correlations will 
contribute more to the delay of the system. For subject 1, there is a hot
spot (central hot spot) in the PMC region of scalp and another in the
frontal region of the 
scalp (i.e.) FCZ (see Fig. 7b). Of the electrodes contributing to the central
hot spot, we do not consider C2 and C6 (as they do not qualify in the surrogate analysis) for further analysis. We
weight the delay obtained for rest of the electrodes contributing to the
central hot spot by $NC^{(l)}(\omega_0)=\frac{C^{(l)}(\omega_0)}{\sum_{j=1}^KC^{(j)}(\omega_0)}$, where $K$
is the number of electrodes contributing to hot spot after qualifying the
surrogate analysis, and $[100 \cdot NC^{(l)}(\omega_0)]$ \% is the contribution of
the $l-$th electrode to the hot spot (it is
easy to check that $\sum_{j=1}^KNC^{(j)}(\omega_0)=1$). Similarly
$NC^{(l)}(\omega_0)$ is calculated for each hot spot. Further, in the calculation of $NC^{(l)}(\omega_0)$
for positive delay, only those electrodes which display positive delay are
considered and similarly for negative delay, only those electrodes which display
negative delay are considered. For subject 1, only one electrode, FCZ
contributes to the frontal hot spot, for which $NC^{(l)}(\omega_0)=1$.  We
have normalised the error-bars of the delays in a similar way. The final
values of delay for both the hot spots are given in Table 1 along with the delay
values obtained by maximising coherence analysis for four other subjects. 
\begin{table}[H]
  \begin{center}
    \begin{tabular}{|c|c|c|c|c|}\hline
      & \multicolumn{4}{c|} {Delay $\delta$ ($ms$) obtained by maximising coherence}\\ \cline{2-5}
      \rb{Subject} & \multicolumn{2}{c|} {Central hot spot} & \multicolumn{2}{c|
}
      {Frontal hot spot}\\ \cline{2-5} 
      & EEG to Muscle & Muscle to EEG & EEG to Muscle
      & Muscle to EEG\\ \hline
      1 & $12.0~\pm~6.6$ & $-16.2~\pm~2.6$ & $25.0~\pm~10.4$ & - \\ \hline
      2 & $14.0~\pm~4.4$ & $-9.0~\pm~4.4$ & $23.0~\pm~7.9$ &$-10.0~\pm~2.3$    \\ \hline
      3 & $11.0~\pm~3.7$ & $-9.0~\pm~3.8$ & $27.3~\pm~12.7$ & - \\ \hline
      4 & $16.0~\pm~7.7$ & $-24.0~\pm~9.0$ & - & - \\ \hline
      5 & $11.34~\pm~6.1$ & $-15.0~\pm~3.8$ & - & - \\ \hline
    \end{tabular}
    \caption{Summary of delay estimation by maximising coherence for five
    essential tremor subjects. Positive and negative signs are introduced in
    maximising coherence analysis just to
    denote the directions of information flow and have got no real life
    significance.}
  \end{center}
\end{table}
Of the five subjects considered three show an additional hot spot in
the frontal region the delay values of which are also given in Table 1 (see third
column). As can be seen there is also a cortico-muscular delay between the frontal hot
spot and the periphery in all three cases. However, this delay is about double
the cortico-muscular delay from the central hot spot. These results indicate
that both the central area most likely corresponding to the primary
sensorimotor cortex and more frontal area are involved in the
generation of essential tremor. While the delays from the central area
(PMC) between 11 and 16 $ms$ are in keeping with a
direct transmission through fast conducting pyramidal pathways \cite{roth} the
cortico-muscular delay from the more frontal area rather indicates another way
of interaction with the periphery. Thus the
method of maximising coherence is capable of distinguishing different types of
cortico-muscular interactions which are relevant to the understanding of the
pathophysiology of essential tremor, for details see \cite{jan1} . The
musculo-cortical delays partly agree but are generally slightly lower than the
somatosensory conduction delays known from routine SEP studies and most likely reflect feed back of the peripheral tremor to the cortex.
\section{Conclusion}
We have used the method of maximising coherence to obtain the time delay between
two series. As a test case, we applied this method to uni- and
bi-directionally coupled R\"ossler attractors. In both cases, the delays
estimated by maximising coherence match well with the expected values,
within the error limits which are obtained by surrogate analysis. For EEG-EMG
time series 
of essential tremor subjects, this method yields a delay in the range of
11 to 16 $ms$ for the primary sensorimotor area and 23-27 $ms$ for the more
frontal area involved in the tremor oscillation whereas the experimentally observed value of the conduction time between primary cortex and muscle is 15 $\pm$ 2 $ms$
\cite{roth}. The larger delay observed for the frontal area may indicate a
different, possibly indirect interaction with the periphery. The musculo-cortical delays (9 to 24 $ms$) are partly in
keeping with the delay observed (around 20 $ms$) in SEP studies. One of the
reasons for the slight deviation of the delay from experimentally
observed value may be as follows: We have employed maximising
coherence method to capture the delay between two series, assuming that
there will be a continuous delayed flow of information from one time series to another. But, it may not be the case in biological data like EEG and EMG. Intermittently the flow may be absent. One way to check out this is to divide the time series into small
sub-portions and calculate the time delay in each portion. But coherence
analysis needs at least a minimum of 30000 data points to decide statistically
whether or not the two time series under study are linearly correlated. So we need to
analyse the nature of information flows using the nonlinear methods
like extended Granger causality \cite{govind}, transfer
entropy \cite{schreiber2}. This issue will be addressed in our future
work. Of course other methods like cross-correlation analysis and time delayed
phase synchronisation analysis \cite{diego} can be used to estimate the delay
between the EEG and EMG 
of the essential tremor subjects. But the problems in using cross-correlation
are addressed in \cite{ml1} and references therein. There exists an analytical
expression for the confidence limit for coherence estimate but this is lacking for
phase synchronisation used in \cite{diego}. That
is the reason why phase synchronisation is not commonly used for
correlation analysis (subsequently for delay analysis in this work) though
this method is expected to be 
superior to coherence analysis \cite{shay}. However comparison of the different methods of
delay estimation is beyond the scope of the current paper.

RBG wishes to acknowledge Dr. Jens Timmer for discussions on surrogate
analysis. Financial support from Deutsche Forschungsgemeinschaft (German
Research Council) is gratefully acknowledged.
 



\end{document}